\pdfoutput=1

\documentclass[runningheads]{llncs}
\usepackage{amssymb}
\usepackage{graphicx}
\usepackage{bm}
\usepackage{subfigure}
\usepackage{multirow}
\usepackage{amsmath}
\usepackage{color,xcolor}
\usepackage{ulem}
\begin{document}
\title{ Controlling False Positive/Negative Rates for Deep-Learning-Based Prostate Cancer Detection on Multiparametric MR images }

\author{Zhe Min$^1$, Fernando J. Bianco$^2$, Qianye Yang$^1$, Rachael Rodell$^{1,3}$, Wen Yan$^{1,4}$, Dean Barratt$^1$, Yipeng Hu$^1$
}
\institute{$^1$ Centre for Medical Image Computing and Wellcome/EPSRC Centre for Interventional \& Surgical Sciences, University College London, London, UK\\
$^2$ Urological Research Network, Miami Lakes, Florida, USA\\
$^3$ Focalyx Technologies, Miami, FL, USA\\
$^4$ City University of Hong Kong, Hong Kong, China\\
\email{z.min@ucl.ac.uk}
}
\authorrunning{Zhe Min \textit{et al.}}
\titlerunning{Controlling FP/NRs for DLB PCa Detection on Multiparametric MRI}
\maketitle    
\begin{abstract}
Prostate cancer (PCa) is one of the leading causes of death for men worldwide. Multi-parametric magnetic resonance (mpMR) imaging has emerged as a non-invasive diagnostic tool for detecting and localising prostate tumours by specialised radiologists. These radiological examinations, for example, for differentiating malignant lesions from benign prostatic hyperplasia in transition zones and for defining the boundaries of clinically significant cancer, remain challenging and highly skill-and-experience-dependent. We first investigate experimental results in developing object detection neural networks that are trained to predict the radiological assessment, using these high-variance labels. We further argue that such a computer-assisted diagnosis (CAD) system needs to have the ability to control the false-positive rate (FPR) or false-negative rate (FNR), in order to be usefully deployed in a clinical workflow, informing clinical decisions without further human intervention. However, training detection networks typically requires a multi-tasking loss, which is not trivial to be adapted for a direct control of FPR/FNR. This work in turn proposes a novel PCa detection network that incorporates a lesion-level cost-sensitive loss and an additional slice-level loss based on a lesion-to-slice mapping function, to manage the lesion- and slice-level costs, respectively. Our experiments based on 290 clinical patients concludes that 1) The lesion-level FNR was effectively reduced from 0.19 to 0.10 and the lesion-level FPR was reduced from 1.03 to 0.66 by changing the lesion-level cost; 2) The slice-level FNR was reduced from 0.19 to 0.00 by taking into account the slice-level cost; (3) Both lesion-level and slice-level FNRs were reduced with lower FP/FPR by changing the lesion-level or slice-level costs, compared with post-training threshold adjustment using networks without the proposed cost-aware training. For the PCa application of interest, the proposed CAD system is capable of substantially reducing FNR with a relatively preserved FPR, therefore is considered suitable for PCa screening applications. 

\end{abstract}

\section{Introduction}
\indent Prostate Cancer (PCa) is one major public health problem for males globally \cite{siegel2020cancer}.  It is estimated that 191,930 cases have been newly diagnosed with PCa and 33,330 associate deaths in the United States in 2020 \cite{siegel2020cancer}.  
Multi-parametric Magnetic Resonance images (mpMR) has potential to play a part in every stage of prostate cancer patient management, including enabling targeted biopsy for early-to-medium stage cancer diagnosis and screening programmes for avoiding unnecessary biopsy \cite{litjens2014computer,wildeboer2020artificial}. However, reading mp-MR requires highly specialised radiologists and, for those experienced, it remains a challenging and arguably tedious task. \\
\indent Automated computer-aided PCa detection not only can help significantly reduce the radiologist's time in examining the volumetric, multi-modality mpMR images, but also provides higher consistency over human interpreters with rivaling human performance at the same time \cite{saha2021end}. Computer-aided diagnosis (CAD) of PCa using mpMR has therefore attracted growing attention and, in particular, modern machine learning methods have been proposed recently for the end-to-end, fully-automated CAD tasks, such as classification, detection and localisation. However, automating PCa detection has to overcome several challenges innate to several imaging and pathology characteristics specific in this application. For example, inherently high inter-patient variance in shape and size among cancerous regions; spatial misalignment between different MR sequences \cite{yu2020deep}; and similar imaging patterns exhibited between the benign prostatic hyperplasia (BPH) and high grade PCa, which subsequently leads to false positives (FPs) \cite{yu2020false,saha2021end}, for both CAD models and human observers, thus their labelling. \\
\indent Scores based on Prostate Imaging and Reporting Data System (PI-RADS) \cite{turkbey2015pirads} and Gleason groups based on biopsy or prostatectomy specimens are examples of radiological and histopathological labels. These two types of labels and their combinations are useful to train a CAD system. Sanford \textit{et al.} utilized a ResNet-based network to assign specific PI-RADS scores to already delineated lesions \cite{sanford2020deep}. Schelb \textit{et al.} compared the clinical performance between PI-RADS and U-Net-based methods for classification and segmentation of suspicious lesions on T2w and diffusion MRI sequences, where the ground-truth is acquired by combined targeted and extended systematic MRI–transrectal US fusion biopsy \cite{schelb2019classification}. While the radiological labels are limited by the challenges discussed above, histopathological labels are also subject to errors and bias in sampling, due to, for example, shift in patient cohort, localisation error in needle biopsy and variance in pathology report. Searching best gold-standard between the two is still an open question and may be beyond the scope of this study. In our work, we use the experienced radiologist PI-RADS scores as our prediction of interest - the training labels. See more details of the data in Section 3. \\
\indent A CAD system for detecting PCa has been considered as a semantic segmentation problem. Many recent PCa segmentation algorithms adopted convolution neural networks (CNNs)  \cite{cao2019joint}. Cao \textit{et al.} has proposed a multiclass CNN called FocalNet to jointly segment the cancerous region and predict the Gleason scores on mpMR images \cite{cao2019joint}. Cao \textit{et al.} adapted the focal loss (FL) to the PCa detection task that predicts the pixel-level lesion probability map on both the Apparent Diffusion Coefficient (ADC) and T2-Weighted (T2w) images, where the training concentrates more on the cancerous or suspicious pixels \cite{cao2019prostate}. In addition, to account for the fact that the lesions may show different size or shapes across imaging modalities, an imaging component selector called selective dense conditional random field is designed to select the best imaging modality where the lesion is observable more clearly \cite{cao2019prostate}. Finally, the predicted probability maps is refined into the lesion segmentation on that selected imaging component \cite{cao2019prostate}. It should be noted that only slices with annotated lesions are included in both the training and validation in \cite{cao2019joint,cao2019prostate}. Yu \textit{et al. } utilised a standalone false positive reduction network with inputs being the detected true positives (TPs) and false positives (FPs) from another U-net-based detection network \cite{yu2020false}. \\
\indent Object detection algorithms have also been proposed for detecting and segmenting PCa from mpMR images, explicitly discriminating between different lesions through instance classification and segmentation. Multiple-staged object detection networks have been shown to have fewer false positives in challenging lesion detecting tasks, compared with segmentation methods such as U-Net \cite{yu2020deep}. Li \textit{et al.} adapted the MaskRCNN to detect the presence of epithelial cells on histological images for predicting Gleason grading \cite{li2018path}, with an addition branch classifying epithelial cell presence in and the MaskRCNN branch classifying, detecting (bounding boxes), and segmenting (into binary masks) the epithelial areas. 
Dai \textit{et al.} investigated the performances of the MaskRCNN to segment prostate gland and the intra-prostatic lesions and reported consistent superior performances over the U-net \cite{dai2020segmentation}. Yu \textit{et al.} also used the MaskRCNN in the PCa detection task, where an additional segmentation branch has been added to the original detection network \cite{yu2020deep}.\\
\indent Two- or multi-stage object detectors have been shown superior performance, compared with the one-stage counterparts \cite{liu2020deep}. However, existing two-stage object detection network in fields of computer vision, such as Mask-RCNN, optimise for overall accuracy, weighting false positive and false negative regions based on their respective prevalence, rather than the associated clinical and societal costs. In this work, we focus on real-world clinical scenarios, in which the CAD system is developed for, for example, assisting population screening or treatment referrals, by alleviating the need for further radiologist examining individual lesions or slices. These clinical applications mandate the developed CAD system to guarantee a low false negative rate and a low false positive rate at lesion or slice levels, in the two respective examples. Instead of thresholding the detection network post-training to achieve the desired sensitivity/specificity at either lesion or slice level, in this study, we aim to answer the research question: With a two-stage object detector, can more desirable FPR or FNR be controlled by changing their costs during training? \\
\indent We explore the plausible answer to this question through formulating and incorporating two cost-sensitive classification losses at the lesion and slice levels respectively, which will give the flexibility of biasing towards reducing FPR or FNR during training. This is not trivial for a detection network training scheme that minimises a multi-tasking loss, as the following technical questions need to be addressed in this work: a) whether a cost-sensitive loss replacing the original instance-level classification loss is effective; b) how slice-level cost can be quantified and subsequently controlled; c) whether changing slice-level cost by the additional slice-level loss is effective; and d) how these two level costs can be combined during training to archive desirable levels of control of FPR/FNR at lesion or slice level, on test data set. \\
\indent Our key contributions of this study are summarised as follows.
(1) We modify the classification loss in the original detection network with the aim of controlling the lesion-level FP/FN for PCa detection. 
(2) We propose a novel slice-level classification loss with the aim of controlling the slice-level FP/FN for PCa detection. We investigate its utility in improving baseline sensitivity with lower FPR by incorporating the classifier into the overall detection network. 
(3) We study the effect of different weighting schemes in the two classifier branches on lesion-level and slice-level FP/FN reduction. 
\section{Methods}
\subsection{Problem definition}
In this work, PCa detection is formulated as an instance segmentation problem. The slices within mpMR images without annotated cancerous regions are regarded as background images. The multiple tasks in PCa detection include: classify whether one proposal region is a lesion or not; regress the coordinates of the bounding box (BB) surrounding the proposal region; segment the mask of the lesion.

\begin{figure*}[t] 
\center{
\includegraphics[width=120mm]{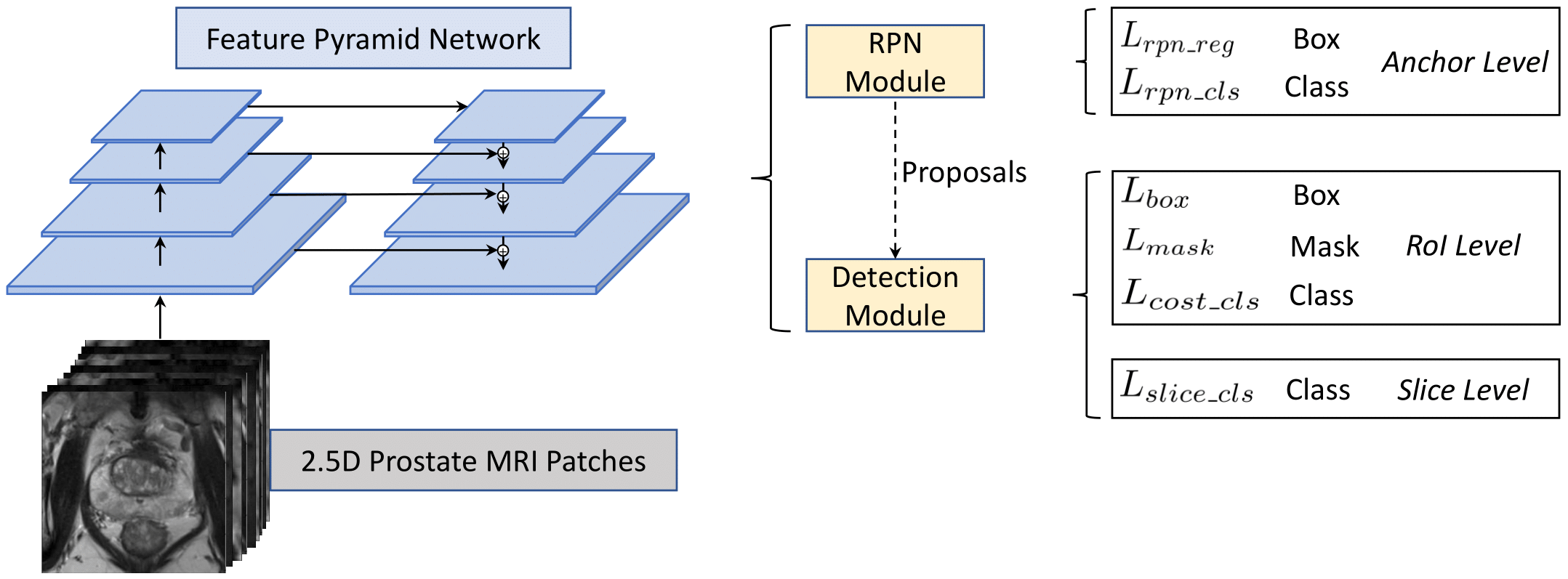}
\caption{Illustration of the overall architecture based on the MaskRCNN. ROI: region of interest, RPN: region proposal network. $L_{rpn\_reg}$: RPN regression loss, $L_{rpn\_cls}$: RPN classification loss, $L_{box}$: the regression loss at the proposal/RoI level, $L_{mask}$: the mask loss at the RoI level, $L_{cost\_cls}$: the lesion-level (i.e., RoI level) cost-sensitive classification loss, $L_{slice\_cls}$: the slice-level cost-sensitive classification loss. }
\label{Overall Architecture Depiction}
}
\end{figure*}

The overall architecture of our proposed CAD system is depicted in Fig. \ref{Overall Architecture Depiction}. 
The network utilizes a Feature Pyramid Network (FPN) backbone on top of the ResNet architecture \cite{he2016deepResidual}, to generate multi-scale features. The extracted features are shared by the following two modules: (a) a region proposal network (RPN) module that generates candidate object bounding boxes \cite{ren2015faster}; (b) a detection module that performs the bounding box regression, classification and the region of interest (RoI) mask prediction on the candidates boxes. 
\subsection{Overall training loss function}
As shown in Fig. \ref{Overall Architecture Depiction}, our multi-task loss consists of the following six terms
\begin{equation}
\label{overall training loss}
L_{total} = L_{rpn\_reg} + L_{rpn\_cls} + L_{box} + L_{mask} + L_{cost\_cls} + L_{slice\_cls}
\end{equation}
where $ L_{rpn\_reg}$ and $ L_{rpn\_cls}$ are the smoothed bounding box regression loss based on ${L^1}$-norm and the cross entropy classification loss, at the anchor level, respectively; $L_{box}$, $ L_{mask}$ and $L_{cost\_cls}$ are the ${L^1}$-norm smoothed bounding box regression loss, the binary cross entropy loss and the weighted cross entropy classification loss, at the RoI level, respectively; and $L_{slice\_cls}$ is the weighted cross entropy classification loss at the slice level. Among all the loss terms in Eq.(\ref{overall training loss}),
$ L_{rpn\_reg}$, $ L_{rpn\_cls}$, $L_{box}$, and $L_{mask} $ are the same as those in the original Mask-RCNN framework.
The rationale of $L_{slice\_cls}$ is to evaluate whether the model can classify the category of a slice being cancerous or not. The inputs to 
$L_{slice\_cls}$ and $L_{slice\_cls}$ are the class probabilities of the proposals being cancerous.

\subsection{Lesion-Level cost-sensitive classification loss}
To control the cost of mis-classification of individual lesions, the lesion-level  (RoI level) cost sensitive classification loss $L_{cost\_cls}(p_i, p_i^{\star})$ is defined as follows
\begin{equation}
\label{cost sensitive loss}
L_{cost\_cls}(p_i, p_i^{\star}) = \underbrace{-\alpha_{lesion} p_i^{\star} \log{p_i} }_{L_{cost\_cls}^{positive}} \underbrace{- \beta_{lesion}(1-p_i^{\star})\log{(1-p_i)}}_{L_{cost\_cls}^{negative}}
\end{equation}
where $p_i^{\star}=1$ if the $i^{th}$ region proposal is positive, $p_i^{\star}=0$ if negative, $p_i\in[0,1]$ is the predicted class probability (by the classification branch in MaskRCNN) of the region proposal $i$ being an cancerous region, $\alpha_{lesion}$ and $ \beta_{lesion}$ are the weights associated with the positive and negative regions. In this study, three different combinations of $\alpha_{lesion}$ and $\beta_{lesion}$ are tested as follows.
(i) $\alpha_{lesion} >1$ and $\beta_{lesion}=1$, during training, the network emphasizes more on regions with positive labels; (ii) $\alpha_{lesion} =1$ and $\beta_{lesion} >1$, the network emphasizes more on the regions with negative labels; (iii) $\alpha_{lesion} =1$ and $\beta_{lesion} =1$, the network weights them equally. In other words, in the above first two cases, the network will penalise more on the (1) false negatives (FNs); (2) false positives (FPs) respectively at the lesion level. 
In the third case, $L_{cost\_cls}(p_i, p_i^{\star})$ degenerates to the binary cross entropy loss when $\alpha_{lesion}=1$ and $\beta_{lesion}=1$. \\
\textbf{Positive Slices} In the slices where there are GT lesions, the training loss associated with that slice is defined as 
\begin{equation}
\begin{split}
     L_{total}= L_{rpn\_reg} + L_{rpn\_cls} + L_{box} + L_{mask} + L_{cost\_cls}.
    \end{split}
\end{equation}
\noindent \textbf{Negative Slices} In the slices where there is no GT lesion, the training loss associated with that slice is defined as 
\begin{equation}
\begin{split}
         L_{total}& = L_{rpn\_cls} + L_{cost\_cls}^{negative}\\
         &=  L_{rpn\_cls} - \beta_{lesion}(1-p_i^{\star})\log{(1-p_i)}.
         \end{split}
\end{equation}

\subsection{Slice-Level cost-sensitive classification loss}
\indent Let us suppose there are $N$ proposal regions or region of interest (ROI) in one slice. The slice-level cost-sensitive classification loss is defined as the weighted cross entropy as follows
\begin{equation}
\label{slice level classification loss}
L_{slice\_cls} = 
\underbrace{- \alpha_{slice} p_{slice}^{\star} \log p_{slice}}_{L_{slice\_cls}^{positive}} - 
\underbrace{\beta_{slice} (1-p_{slice}^{\star} ) \log(1-p_{slice})}_{L_{slice\_cls}^{negative}},
\end{equation}
where $p_{slice}^{\star}\in\{0,1\}$ and $p_{slice}\in[0,1]$ is given by
\begin{equation}
 p_{slice}^{\star} = max(p^{\star}_1,..., p^{\star}_N),
\end{equation}
\begin{equation}
p_{slice} = max(p_1,..., p_N),
\end{equation}
where $p_{i}^{\star}$ and $p_{i}$ are the GT and predicted probability that $i^{th}$ region being cancerous. More specifically, 
$p_{slice}^{\star}=1$ indicates that there is at least one cancerous region in the interested slice, $p_{slice}$ is the largest predicted probability of one detected region being cancerous. The rational behind the lesion-to-slice mapping function, for computing $p_{slice}^{\star}$ and $p_{slice}$, is that (1) for GT labels, one slice is considered to be a `cancerous' slice if there exists at least one `positive' region (i.e., $p_i^{\star}=1$ for at least one $i$); (2) for predictions, the  probability of one slice being `cancerous' is the largest predicted probability of one detected region being 'positive' in the interested slice. Like the function of $\alpha_{lesion}$ and $\beta_{lesion}$ in $L_{cost\_cls}(p_i, p_i^{\star})$, $\alpha_{slice}$ and $\beta_{slice}$ weight the loss $L_{slice\_cls}$ in an adversarial manner: Whilst $\alpha_{slice}>1$ and $\beta_{slice}=1$, the network penalises FNs more heavily at the slice level, $\alpha_{slice}=1$ and $\beta_{slice}>1$, the network penalises FPs more.  \\
\textbf{Positive Slices} In the slices where there are GT lesions, the overall training loss remains $L_{total}$, defined in Eq.(\ref{overall training loss}), and can be expanded as follows
\begin{equation}
    L_{total} = L_{rpn\_reg} + L_{rpn\_cls} + L_{box} + L_{mask} + L_{cost\_cls} - \alpha_{slice} p_{slice}^{\star} \log p_{slice}.
\end{equation}
\textbf{Negative Slices} In the slices where there is no GT lesion, the overall training loss is therefore given by
\begin{equation}
\begin{split}
    L_{total} &= L_{rpn\_cls} + L_{cost\_cls}^{negative} + L_{slice\_cls}^{negative}\\
    &= L_{rpn\_cls}  - \beta_{lesion}(1-p_i^{\star})\log{(1-p_i)}- \beta_{slice} (1-p_{slice}^{\star} ) \log(1-p_{slice}).
    \end{split}
\end{equation}
where only the classification losses at the anchor, lesion/region, and slice levels are included.

\section{Experiments and Evaluation}

\subsection{Data set and implementation details}
Our data sets consist of 290 clinical prostate cancer patients with approved Institutional Review Board (IRB) protocol. The ground-truth labels (including cancerous masks) have been acquired based on the Prostate Imaging Reporting and Data System (PI-RADS) scores reported by radiologists with more than 15 years of experience. PIRADS $\geq 3$ annotated lesions are regarded as clinically significant and are considered positive in this work. The ratios of number of patients in the training, validation and test sets are 8:1:1. The inputs to our proposed detection include the T2-Weighted (T2w), the Apparent Diffusion Coefficient (ADC), and the Diffusion-Weighted Images (DWI) b-2000 images. 
ADC and DWI b-2000 images were spatially aligned with corresponding T2w images using the rigid transformation based on the coordinate information stored in the imaging files. All slices were cropped from the center to be 160$\times$160 and 
the intensity values were normalized to [0,1]. Our networks were constructed with 2D convolutional layers, with a so-called 2.5D input bundle which concatenated two neighboring slices for each of the T2, ADC and DWI b-2000 image slices at the slice of interest, i.e. resulting in a nine-channel input as denoted in Fig. \ref{Overall Architecture Depiction}.

The proposed method was implemented with the TensorFlow framework. 
Each network was trained for 100 epochs with the stochastic gradient descent (SGD) optimizer and the initial learning rate was set to be 0.001. Random affine transformations were applied for data augmentation during training. If not otherwise specified, the parameter \textsf{threshold}\footnote{We use \textsf{threshold} to denote parameter DETECTION$\_$MIN$\_$CONFIDENCE in the original MaskRCNN codes, for brevity.} was set to 0.7 and the maximum number of lesions in one slice being 6 was configured at both the training and test stages. 

\subsection{Evaluation metrics} 
We evaluate the methods with descriptive statistics at both the lesion and slice levels. The slice-level false positive rate (FPR) and false negative rate (FNR) are defined as follows
$\textrm{FPR} = \frac{\textrm{FP}}{\textrm{FP}+ \textrm{TN}} = 1 - \textrm{specificity}
$, $
\textrm{FNR} = \frac{\textrm{FN}}{\textrm{FN}+\textrm{TP}} = 1- \textrm{sensitivity}
$, $
    \textrm{ACC} = \frac{\textrm{TP}+ \textrm{TN}}{\textrm{TP}+
    \textrm{TN}+\textrm{FP}+\textrm{FN}}
$, 
where FP, TN FN and TP are numbers of false positive, true negative, false negative and true positive cases, respectively. It is noteworthy that the above definitions are defined and used at the slice level. At the lesion level, only the definition of FNR remains valid. Instead, we compute the mean FP per slice.  \\
\indent \textbf{At the lesion level}, a TP prediction requires the GT lesion has an Intersection of Union (IoU) greater than or equal to 0.2, between the GT bounding box (BB) and any predicted BB. A FP prediction means IoUs are smaller than 0.2 (including no overlap) between the predicted BB and all GT BBs. A GT lesion that has no TP prediction is counted as a FN. TN is not defined at the lesion level. \\
\indent \textbf{At the slice level}, one slice with at least one GT annotated lesion mask is considered as a TP if there is any detected region at that slice. If there is no detection on the slices with GT lesion masks, the slice is counted as a FN. A TN slice means no lesion found in both prediction and GT. Any positive lesion predicted on a slice that has no GT lesion leads to a FP slice.
\begin{table}[t]
    \centering
    \begin{tabular}{c|c|c|c}
    &$\alpha_{lesion}/\beta_{lesion}=1$& $\alpha_{lesion}=3$, $\beta_{lesion}=1$ &  $\alpha_{lesion}=1$, $\beta_{lesion}=3$    \\
    \hline
        Lesion-level FP& 1.0327 & 2.0218 & $\bm{0.6567}$\\
        \hline
    Lesion-level FNR& 0.1941& $\bm{0.1013}$ &   0.4118    \\
                    \hline
 Slice-level FPR&0.5878  &0.8434  &  $\bm{0.5049}$   \\
        \hline
   Slice-level FNR&0.0097 & $\bm{0.0028}$  &  0.0736    \\
             \hline
    ACC&  0.5744 &  0.3924 & $\bm{0.6161}$
    \end{tabular}
    \caption{The false positive rate (FPR) and false negative rate (FNR) on the test data sets, where $L_{cost\_cls}$ was used in the training process. With $\alpha_{lesion}=3$, $\beta_{lesion}=1$ in Eq.(\ref{cost sensitive loss}), both the lesion-level and slice-level FNRs were considerably reduced, compared to the case where $\alpha_{lesion}=1,\beta_{lesion}=1$. With $\alpha_{lesion}=1$, $\beta_{lesion}=3$ in Eq.(\ref{cost sensitive loss}), the lesion-level FPs and slice-level FPRs were lower, compared to the case where $\alpha_{lesion}=1,\beta_{lesion}=1$.  }
      \label{results that incorporate the weights at the lesion level}
\end{table}
\begin{table}[t]
    \centering
    \begin{tabular}{c|c|c|c}
    &$\alpha_{slice}=1$, $\beta_{slice}=1$& $\alpha_{slice}=3$, $\beta_{slice}=1$ &  $\alpha_{slice}=1$, $\beta_{slice}=3$    \\
    \hline
Lesion-level FP &$\bm{1.7202}$  &1.9493  & 1.7965  \\
        \hline 
Lesion-level FNR &0.1190 & $\bm{0.0970}$ &  0.1232     \\
                    \hline
   Slice-level FPR & 0.8505 & $\bm{0.8234}$ &  0.8277   \\
        \hline
   Slice-level FNR &  $\bm{0.0000}$& $\bm{0.0000}$  &0.0014\\
       \hline
    ACC & 0.3882  & $\bm{0.4076}$  &0.4041 \\
    \end{tabular}
    \caption{The false positive rate (FPR) and false negative rate (FNR) on the test data sets where $L_{cost\_cls}$($\alpha_{lesion}/\beta_{lesion}=1$) and $L_{slice\_cls}$ were incorporated into the training. With $\alpha_{slice}=3,\alpha_{slice}=1$ in Eq.(\ref{slice level classification loss}): (a) the lesion-level FNR was reduced; (b) the slice-level FNR remained close to zero with reduced slice-level FPR, compared to the case where $\alpha_{slice}=1,\beta_{slice}=1$. With $\alpha_{slice}=1,\beta_{slice}=3$ in Eq.(\ref{slice level classification loss}),  the slice-level FPR was reduced, compared to the case where $\alpha_{slice}=1,\beta_{slice}=1$. 
    }
    \label{tab:my_label2}
\end{table}
\begin{table}[t]
    \centering
    \begin{tabular}{c|c|c|c}
     &$\alpha=1$,$\beta=1$ & $\alpha=3$ $\beta=1$ &  $\alpha=1$, $\beta=3$
   \\ 
Lesion-level FP  & 1.7202& 2.3827 &$\bm{1.0982}$\\
        \hline 
Lesion-level FNR  & 0.1190 &  $\bm{0.0734}$ & 0.2262   \\
    \hline
    Slice-level FPR &0.8505 &0.9220  &$\bm{0.6576}$     \\
        \hline
 Slice-level FNR &  $\bm{0.0000}$& $\bm{0.0000}$ & 0.0014  \\
             \hline
    ACC & 0.3882  & 0.3367  &$\bm{0.5265}$ \\
    \end{tabular}
    \caption{The false positive rate (FPR) and false negative rate (FNR) on the test data sets where $L_{cost\_cls}$ and $L_{slice\_cls}$ were incorporated. With $\alpha=3,\beta=1$ in Eq.(\ref{cost sensitive loss}) and Eq.(\ref{slice level classification loss}), (a) the lesion-level FNR was reduced; (b) the slice-level FNR remained to be 0, compared to the case where $\alpha=1,\beta=1$. With $\alpha=1,\beta=3$,  the lesion-level FP and slice-level FPR were reduced, compared to those where $\alpha=1,\beta=1$. 
    }
    \label{Reducing the false negative rate with the slice level.}
\end{table}
\begin{figure*}[t] 
\center{
\includegraphics[width=120mm]{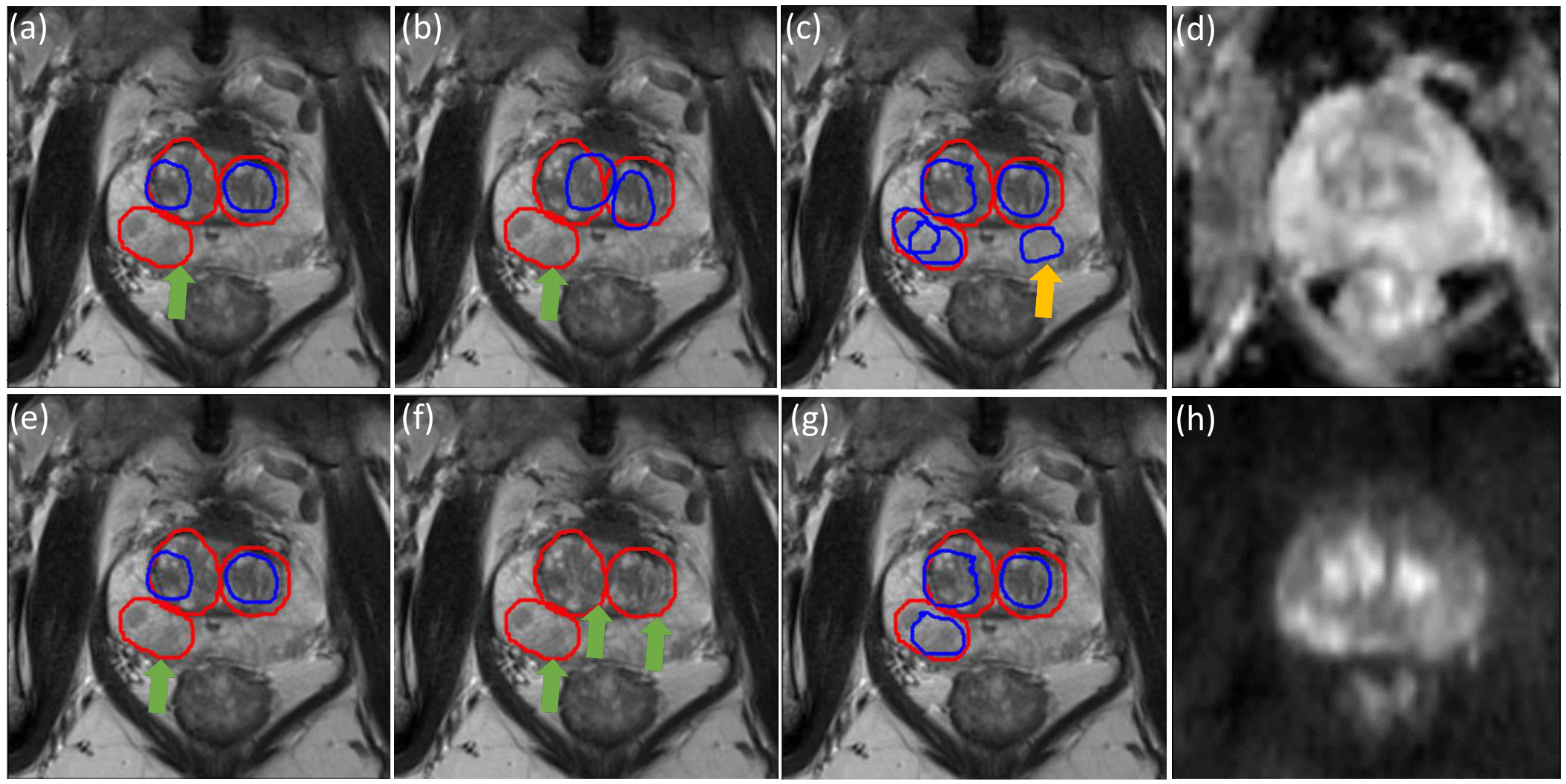}
\vspace{1mm} 
\caption{
In all figures in this paper, (1) the red circles denote the ground-truth (GT) lesion region, while the blue circles denote the predicted regions of interest; (2) a false positive (FP) predicted detection is denoted with the yellow arrow, while another false negative (FN) lesion is denoted with the green arrow. 
In this study, only the lesion-level classification loss $L_{cost\_cls}$ in the training process.  All example sub-figures shown here correspond to the performances on one same slice in the test data set. In the first row,  \textsf{threshold}=0.7 while \textsf{threshold}=0.95 in the second row.
The first three columns from the left show the detected results with only $L_{cost\_cls}$ incorporated into the training loss. 
(a,e) $\alpha_{lesion}=1, \beta_{lesion}=1$; (b,f) $\alpha_{lesion}=1, \beta_{lesion}=3$; (c,g) $\alpha_{lesion}=3, \beta_{lesion}=1$. (d) Apparent Diffusion Coefficient (ADC) image; (h) Diffusion-Weighted Images (DWI) b-2000 image. 
}
\label{Reduce FNR with the modified loss at the lesion level}}
\end{figure*}
\section{Results}
\subsection{Adjusting mis-classification cost at lesion-level}
In this experiment, we study the impact on the lesion-level and slice-level performances, due to different $L_{cost\_cls}$ in the training, while the slice-level loss $L_{slice\_cls}$ in Eq.(\ref{slice level classification loss}) is not included. 
More specifically, we compare the original MaskRCNN (i.e., $\alpha_{lesion}=1, \beta_{lesion}=1$ in Eq.(\ref{cost sensitive loss})), with our proposed two variants where $\alpha_{lesion}=3, \beta_{lesion}=1$ and $\alpha_{lesion}=1, \beta_{lesion}=3$. \\
\indent Table \ref{results that incorporate the weights at the lesion level} summarises the comparative results with the case where $\alpha_{lesion}=1,\beta_{lesion}=1$. With $\alpha_{lesion}=3$, $\beta_{lesion}=1$, the lesion-level and slice-levels FNRs were reduced from 0.1941 to 0.1013, from 0.0097 to 0.0028, respectively. With $\alpha_{lesion}=1$, $\beta_{lesion}=3$, the lesion-level FP was reduced from 1.0327 to 0.6567 while the slice-level FPR was reduced from 0.5878 to 0.5049.  
\begin{figure*}[t] 
\center{
\includegraphics[width=120mm]{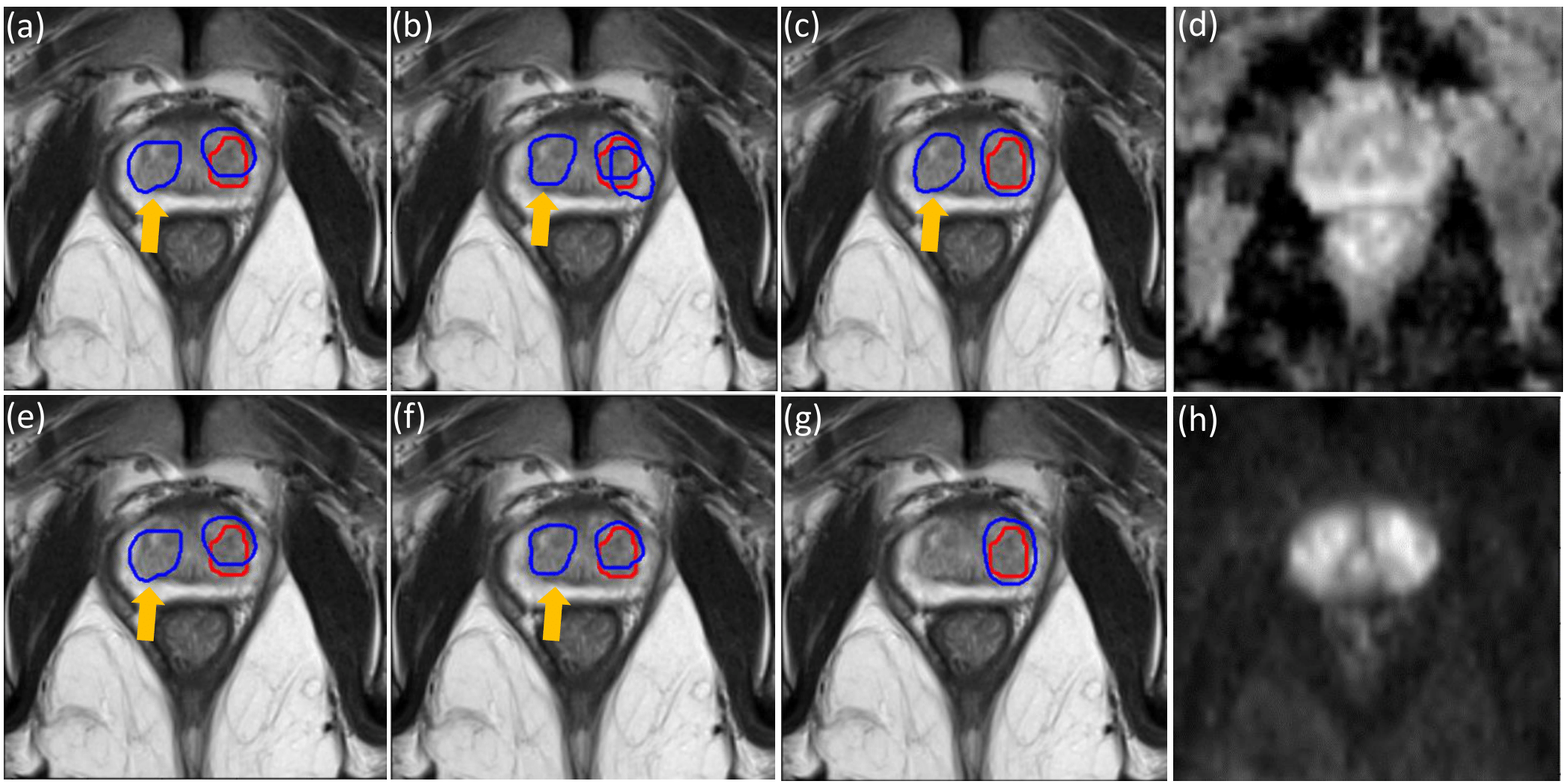}
\caption{This figure demonstrates the reduction of the lesion-level FPs by changing the lesion-level classification cost $L_{cost\_cls}$. 
The same setting in training was adopted as that in Fig. \ref{Reduce FNR with the modified loss at the lesion level}, and all example sub-figures shown here correspond to the performances on one same slice in the test data set (but a different slice with that in Fig. \ref{Reduce FNR with the modified loss at the lesion level}.
In the first row, \textsf{threshold}=0.7 while \textsf{threshold}=0.95 in the second row. The weighting schemes are summarised as follows: 
(a,e) $\alpha_{lesion}=1, \beta_{lesion}=1$; (b,f) $\alpha_{lesion}=3, \beta_{lesion}=1$; (c,g) $\alpha_{lesion}=1,\beta_{lesion}=3$. (d) ADC image; (h) DWI b-2000 image.    }
\label{Reduce FPR with the modified loss at the lesion level}}
\end{figure*}
\begin{figure*}[t] 
\center{
\includegraphics[width=120mm]{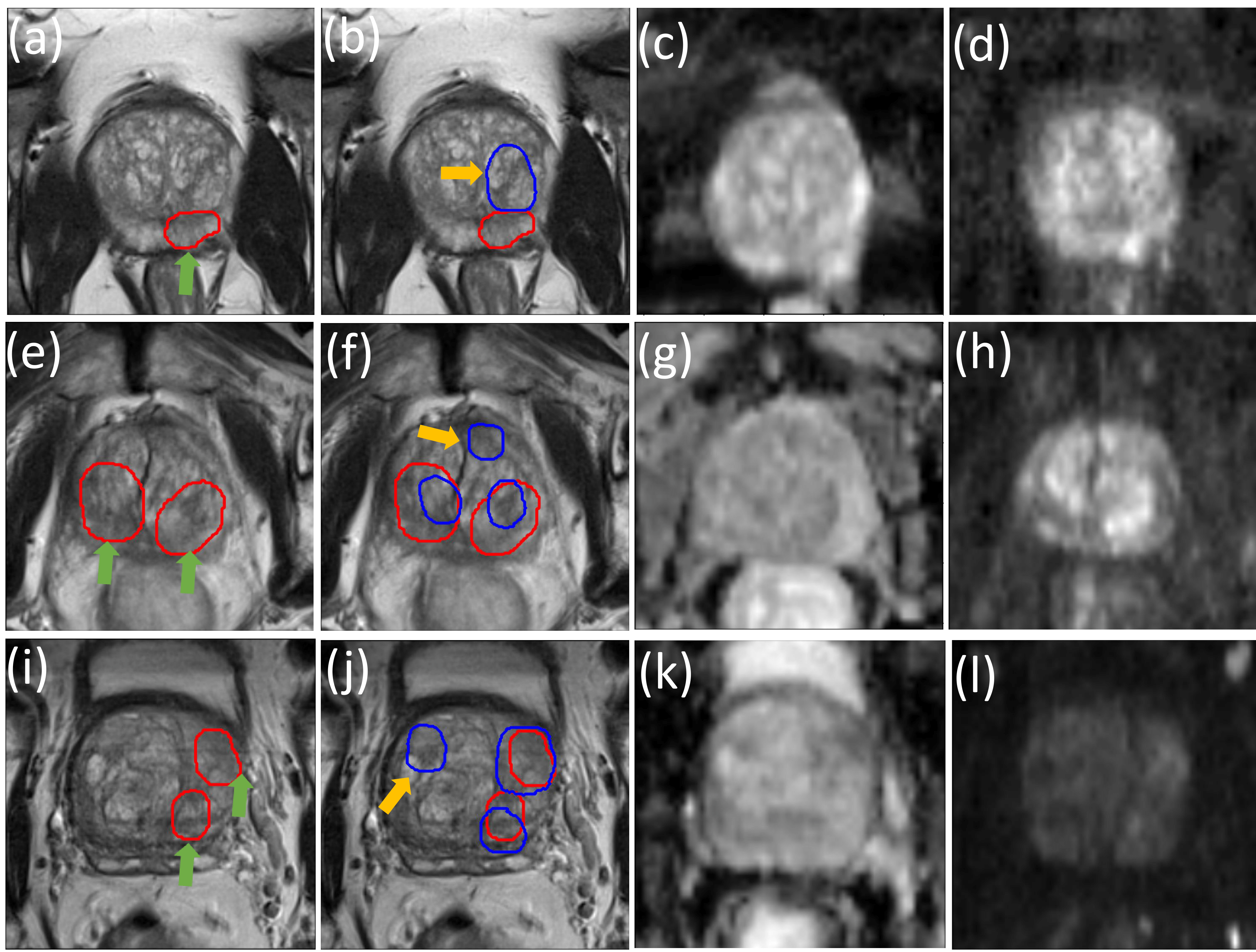}
\caption{
This figure demonstrates that both the lesion-level and slice-level FNs were reduced by incorporating $L_{slice\_cls}$ into the training process. In all the ablation examples presented in this figure, $\textsf{threshold}=0.95$. 
Example sub-figures shown here in the same row correspond to the same slice in the test data set. (a,e,i) depicts the detection results with only the lesion-level classification loss $L_{cost\_cls}$ incorporated, where $\alpha_{lesion},\beta_{lesion}$ vary. 
(b,f,j) depicts the detection results with both $L_{cost\_cls}$ and $L_{slice\_cls}$ utilized in the training. 
The weighting schemes in this ablation study are as follows. (a) $\alpha_{lesion}=1, \beta_{lesion}=1$; 
(b) $\alpha=1,\beta=1$; 
(e) $\alpha_{lesion}=1, \beta_{lesion}=3$; (f) 
$\alpha=1,\beta=3$;
(i) $\alpha_{lesion}=3,\beta_{lesion}=1$; (j)
$\alpha=3,\beta=1$. (c,g,k) ADC images; (d,h,l) DWI b-2000 images.
}
\label{Reduce FNR with the modified loss at the slice level}}
\end{figure*}

\indent  Fig. \ref{Reduce FNR with the modified loss at the lesion level} shows the examples where the FNs were reduced with $\alpha_{lesion}=3$, $\beta_{lesion}=1$, by comparing Fig. \ref{Reduce FNR with the modified loss at the lesion level} (c) with Fig. \ref{Reduce FNR with the modified loss at the lesion level}. (a,b), and comparing Fig. \ref{Reduce FNR with the modified loss at the lesion level} (g) with Fig. \ref{Reduce FNR with the modified loss at the lesion level} (e,f).
By comparing Fig. \ref{Reduce FNR with the modified loss at the lesion level} (g) with Fig. \ref{Reduce FNR with the modified loss at the lesion level} (c), the FP was reduced with a higher \textsf{threshold}. In contrast, more FNs can be found with larger \textsf{threshold} and $\alpha_{lesion}=1,\beta_{lesion}=3$ by comparing Fig. \ref{Reduce FNR with the modified loss at the lesion level} (f) with Fig. \ref{Reduce FNR with the modified loss at the lesion level} (b).\\
\indent Fig. \ref{Reduce FPR with the modified loss at the lesion level} shows the example where the FPs were avoided/reduced with $\alpha_{lesion}=1$, $\beta_{lesion}=3$, by comparing Fig. \ref{Reduce FPR with the modified loss at the lesion level} (g) with Fig. \ref{Reduce FPR with the modified loss at the lesion level} (a,b,c,e,f).  In the first row in Fig. \ref{Reduce FPR with the modified loss at the lesion level} (c), with relatively lower value of the parameter \textsf{threshold}, the FP still exists with $\alpha_{lesion}=1,\beta_{lesion}=3$. In contrast, by comparing Fig. \ref{Reduce FPR with the modified loss at the lesion level} (g) with Fig. \ref{Reduce FPR with the modified loss at the lesion level} (e,f), with larger value of the parameter \textsf{threshold}, the FP was avoided as shown in Fig. \ref{Reduce FPR with the modified loss at the lesion level} (g). 
\subsection{Adjusting mis-classification cost at slice-level} 
\indent 
In this experiment, we study the effect of incorporating and changing $L_{slice\_cls}$ in the training loss whereas the weighting in $L_{cost\_cls}$ was fixed as $\alpha_{lesion}=1, \beta_{lesion}=1$.
Table \ref{tab:my_label2} includes the quantitative results with different settings of $\alpha_{slice}, \beta_{slice}$: (1)
$\alpha_{slice}=1,\beta_{slice}=1$; (2) $\alpha_{slice}=3,\beta_{slice}=1$; (3) $\alpha_{slice}=1,\beta_{slice}=3$.
With $\alpha_{slice}=3, \beta_{slice}=1$, (a) the lesion-level FNR was reduced from 0.1190 to 0.0970; (b) the slice-level FNR remained to be 0.0000 while the slice-level FPR was also reduced from 0.8505 to 0.8234, compared to the case where $\alpha=1,\beta=1$. 
With $\alpha_{slice}=1, \beta_{slice}=3$, (a) the FPR was reduced from 0.8505 to 0.8277; (b) the lesion-level FP was increased from 1.7202 to 1.7965, compared to the case where $\alpha=1,\beta=1$.\\
\indent By comparing the second column in Table \ref{results that incorporate the weights at the lesion level} and the second column in Table \ref{tab:my_label2}, we can find that the lesion-level and slice-level FNRs were reduced from 0.1941 to 0.1190 and from 0.0097 to 0, respectively. 
Comparing the third column in Table \ref{results that incorporate the weights at the lesion level} and the third column in Table \ref{tab:my_label2}, finds that lesion-level and slice-level FNRs were reduced from 0.1013 to 0.0970 and from 0.0028 to 0.0000 respectively while (1) the lesion-level FP was reduced from 2.0218 to 1.9493; (2) the slice-level FPR was reduced from 0.8434 to 0.8234. The improvements in both FPRs and FNRs, by incorporating and further changing the slice-level cost, indicate the benefits and the significance of using the slice-level cost-sensitive classification loss.

\subsection{Adjusting mis-classification cost at both levels} 
In this experiment, we study the effect of changing both $L_{cost\_cls}$ and $L_{slice\_cls}$ on the performance by varying $\alpha$ and $\beta$. Table \ref{Reducing the false negative rate with the slice level.} shows the corresponding results with three different settings of $\alpha$ and $\beta$: (a) $\alpha_{lesion/slice}=1, \beta_{lesion/slice}=1$; (b) $\alpha_{lesion/slice}=3, \beta_{lesion/slice}=1$; (c) $\alpha_{lesion/slice}=1, \beta_{lesion/slice}=3$. With $\alpha=3, \beta=1$, compared to the case where $\alpha=1,\beta=1$, (a) the lesion-level FNR was reduced from 0.1190 to 0.0734; (b) the slice-level FNR remained to be 0. With $\alpha=1, \beta=3$, compared to the case where $\alpha=1,\beta=1$, (a) the lesion-level FP was reduced from 1.7202 to 1.0982; (b) the slice-level FPR was reduced from 0.8505 to 0.6576. \\
 \indent 
 By comparing the corresponding results in the same columns in Table \ref{Reducing the false negative rate with the slice level.} with those in Table \ref{results that incorporate the weights at the lesion level} respectively, both the lesion-level and slice-level FNRs were substantially reduced by incorporating the slice-level classification loss $L_{slice\_cls}$ into training.
 By comparing corresponding results in the third column in Table \ref{Reducing the false negative rate with the slice level.} with those in Table \ref{tab:my_label2}, (1) the lesion-level FNR was reduced
 from 0.0970 to 0.0734; (2) the slice-level FNR remained to be 0.  By comparing corresponding results in the last column in Table \ref{Reducing the false negative rate with the slice level.} with those in Table \ref{tab:my_label2}, it becomes clear that (1) the lesion-level FP was reduced from 1.7965 to 1.0982; (2) the slice-level FPR was reduced from 0.8277 to 0.6576. 
 \\
\indent Fig. \ref{Reduce FNR with the modified loss at the slice level} includes the three ablation examples where the slice-level FNs were reduced by incorporating the slice-level classification loss $L_{slice\_cls}$ into training.
Three different slices are utilized to demonstrate the improvements in the three different rows in Fig. \ref{Reduce FNR with the modified loss at the slice level}.
Comparing Fig. \ref{Reduce FNR with the modified loss at the slice level} (b) with Fig. \ref{Reduce FNR with the modified loss at the slice level} (a), shows that the slice-level FN was reduced with the sacrifice of one more lesion-level FP.  By comparing Fig. \ref{Reduce FNR with the modified loss at the slice level} (f) with Fig. \ref{Reduce FNR with the modified loss at the slice level} (e), we find that both lesion-level and slice-level FNs were reduced with one more lesion level FP. By comparing Fig. \ref{Reduce FNR with the modified loss at the slice level} (j) with Fig. \ref{Reduce FNR with the modified loss at the slice level} (i), we find that both lesion-level and slice-level FNs were reduced with the sacrifice of one more lesion-level FP. 

\subsection{Results analysis}
It should be noted that all the terms in the loss are weighted equally in this work. The effects of different weighting factors associated with different sub-tasks will be explored in the future. In addition, a wider range of $\alpha$ and $\beta$ will be tested to find their optimal values. 
In this section, we quantitatively analyse the impact of changing the training-time cost-sensitive losses, compared with those where the threshold parameter was adjusted post-training.  For brevity, in what follows, we use (1) $\alpha_{lesion}, \beta_{lesion}$ to refer to the case where only the cost-sensitive loss $L_{cost\_cls}$ was used in training; (2)
$\alpha_{slice},\beta_{slice}$ to refer to the case where $\alpha_{lesion}=1,\beta_{lesion}=1$ while the cost-sensitive slice-level loss $L_{slice\_cls}$ was also utilized in training, and the weights may vary; (3) $\alpha,\beta$ to refer to the case where both $L_{cost\_cls}$ and $L_{slice\_cls}$ were used in training, and the weights in the both losses can change. 

We further group the interesting conclusions into positive and negative results, indicating the resulting impact difference to our specific PCa application. These, however, may not generalise to other clinical applications that adopt the same proposed cost-adjusting strategies.
\subsubsection{Positive Results}
\begin{enumerate}
\item With $\alpha_{lesion}=1$, $\beta_{lesion}=1$, by adjusting the post-training threshold, the lesion-level FNR was reduced to 0.1131 with the lesion-level FP being $\bm{5.9758}$.  
In contrast, (1) with $\alpha_{lesion}=3, \beta_{lesion}=1$, the lesion-level FNR was 0.1013 
while the FP was $\bm{2.0218}$; 
(2) with $\alpha_{slice}=1, \beta_{slice}=1$, the lesion-level FNR was 0.1190 while the FP was $\bm{1.7202}$; 
(3) with $\alpha_{slice}=3, \beta_{slice}=1$, the lesion-level FNR was 0.0970 
with the FP was $\bm{1.9493}$. To summarize, by choosing the appropriate loss during training, a considerable lower FP value can be achieved with comparable or reduced lesion-level FNs, compared to those from changing the threshold. 
\item With $\alpha_{lesion}=1, \beta_{lesion}=1$, by adjusting the threshold, the slice-level FNR was reduced to be 0.0042 with the FPR being $\bm{0.6972}$. In contrast, with $\alpha=1,\beta=3$, the slice-level FNR was 0.0014 while the FPR was $\bm{0.6576}$.    
\item With $\alpha_{slice}=1, \beta_{slice}=1$, by adjusting the threshold, the lesion-level FNR was reduced to 0.0987 while the FP was $\bm{2.0530}$. 
In contrast, with $\alpha_{slice}=3, \beta_{slice}=1$, the lesion-level FNR and FP were 0.0970 and $\bm{1.9493}$, respectively.
\item With $\alpha_{slice}=3,\beta_{slice}=1$, compared to the case where $\alpha_{slice}=1,\beta_{slice}=1$,  the slice-level FPR was reduced to 0.8234 while the FNR remained to be 0.     
\item With $\alpha=1, \beta=1$, by adjusting the threshold, the lesion-level FNR was reduced to be 0.0734 
while the lesion-level FP was $\bm{5.4910}$. 
In contrast, with $\alpha=3,\beta=1$, 
the lesion-level FP was $\bm{2.3827}$ while the FNR was 0.0734. 
\item With $\alpha=1,\beta=1$, the slice-level FNR was reduced to be 0.014 with the slice-level FPR being $\bm{0.7161}$. In contrast, with $\alpha=1,\beta=3$, the slice FNR and the slice-level FPR were 0.0014 and $\bm{0.6576}$, respectively.
\end{enumerate}
Comparing results in 1 and 5, at the lesion level, shows that the added FPs can be reduced in order to achieve a lower FNR by simply adding the classification loss at the slice level. The above results demonstrate the significant advantage of incorporating the cost-sensitive classification loss in reducing the lesion-level and slice-level FNRs.

\subsubsection{Negative Results}
\begin{enumerate}
    \item With $\alpha=1, \beta=1$, by adjusting the threshold, the slice-level FNR was reduced to be 0 with the FPR being 0.9166, which is smaller than 0.9220 where $\alpha=3,\beta=1$.
    \item With $\alpha_{lesion}=1, \beta_{lesion}=1$, by adjusting the threshold, the lesion-level FP was reduced to be $\bm{0.4774}$ with the FNR being $\bm{0.3662}$. These two values are smaller than those where $\alpha_{lesion}=1, \beta_{lesion}=3$, respectively.  
    \item At the slice level where $\alpha=1,\beta=1$, the FNR was reduced to be 0.014 with FPR being $\bm{0.7161}$. In contrast, in the case where $\alpha=1,\beta=3$, FNR was 0.0014 with FPR being $\bm{0.8277}$.
\end{enumerate}
In the training data set, the class imbalance problem was present where much more background objects/slices exist. Interestingly, this is the reason we believe that the so-called negative results originated in this application, in which, a greater weighting towards the majority class(es) would further reduce the biased (usually lower) prediction performance on the minority class(es), although the associated costs may have been correctly minimised. Further analysis for this phenomenon between prediction performance with- and without considering costs might warrant further investigation.

\section{Conclusions}
In this study, we explore the feasibility of controlling the false positives/negatives at the lesion or slice level during training, together with an in-depth analysis of the associated advantages and disadvantages. We conclude the quantitative results obtained from the clinical patient data set as follows: 1) Incorporating the proposed cost-sensitive classification losses at either lesion or slice level (or both) demonstrates the expected flexibility of controlling the false positive rate (FPR) and false negative rate (FNR); and
2) Incorporating the proposed cost-aware losses was able to reduce the FNRs while maintaining or further reducing the FPRs, which can be particularly useful for real-world clinical applications such as population screening for prostate cancer. 

\section*{Acknowledgements}
This work is supported by the Wellcome/EPSRC Centre for Interventional and Surgical Sciences (203145Z/16/Z). This work was supported by the International Alliance for Cancer Early Detection, a partnership between Cancer Research UK [C28070/A30912; C73666/A31378], Canary Center at Stanford University, the University of Cambridge, OHSU Knight Cancer Institute, University College London and the University of Manchester.

\bibliographystyle{splncs03}   

\bibliography{miccai2021}

\end{document}